\documentclass[aps, reprint, floatfix]{revtex4-1}
\usepackage{graphicx}
\usepackage{bm}
\usepackage[utf8]{inputenc}
\usepackage[T1]{fontenc}

\begin{document}

\preprint{AIP/123-QED}

\title[]{Carboxyl and Amine Functionalized Carboranethiol SAMs on 
Au(111) :\\ A Dispersion Corrected Density Functional Theory Study}

\author{Merve Yortanl{\i}}
\author{Ersen Mete}
\email[Corresponding author : ]{emete@balikesir.edu.tr}

\affiliation{Department of Physics, Bal{\i}kesir University, Bal{\i}kesir 10145, 
Turkey}

\date{\today}

\begin{abstract}
The morphological and electronic properties of isolated and monolayer phases 
of carboxyl- and amine-functionalized carboranethiols on unreconstructed 
Au(111) were determined using density functional theory calculations by 
including van der Waals interactions. The groups are effective in the assembly 
of pristine adlayers on gold and also offer functionality when exposed at the 
SAM-environment interface. Anisotropy brought by both functional groups 
increases tilting of carboranethiols relative to the surface normal 
and absolute values of the dissociative chemisorption energies. Positional 
isomerization and the functional groups modify the molecular dipole moments 
which influnce the adsorption characteristics. Even though carboxylic acid 
and amine groups have different chemical properties, they have similar effects 
on the adsoprtion characteristics of carboranethiols. Dense packing favors 
intermolecular interactions which gives a stronger binding relative to isolated 
adsorption. The carboranethiols with the functional groups can be arranged in 
the same lateral direction or in a dimer conformation with molecues facing 
each other. Carboxyl and amine groups offer functionalization of 
cabranethiol SAMs and in conjuction with positional isomerization shift 
tunability ranges of the work function of gold to higher energies.
\end{abstract}



\maketitle

\section{Introduction}

Thiol based (R-S-H) self-assembled monolayers (SAM) are promising for 
applications in different fields such as molecular electronics, nanotechnology, 
surface wetting and molecular recognition.\cite{Kumar, Bishop, Schon, Fendler, 
Chaki, Gooding, Love, Frasconi, Vericat, Mandler} The characteristics of SAMs 
are related to coated molecule on the surface as well as to the coating density. 
The molecular density of SAMs is effective in inter-molecular and molecule 
surface interactions and plays a crucial role in the film's structural 
and electronic properties. For instance, work function of metal electrodes 
used in the design of organic electronic devices can be modified by adjusting 
dipole moments of different thiol SAMs attached to those electrodes. Thiol 
SAMs provide a great advantage in many applications, and therefore, has 
been the subject of many experimental\cite{De2005,Base2005,Balaz2007, 
Gozlan,Base2008,Crispin,Zehner,Ishii,Hohman2009,Lee2011,Lubben2011,Scholz2011,
Base2010,Base2012,Albayrak2013,Albayrak2014,Kim2014,Serino2017,Yavuz2017,
Foerster2019} and theoretical\cite{Gronbeck,Rusu1,Rusu2,Romaner,Sushko2009, 
Otalvaro2012,Osella2014,Mete2016,Cornil,Hladik2019} studies. Moreover, 
additional functional groups were shown to be effective in modification of the 
work function of gold substrates. Lee~\textit{et al.} demonstrated the 
tunability of the work function by amine- and carboxylic acid-functionalized 
alkanethiol SAMs.\cite{Lee2011} 

Dicarbacloso-dodecaborane thiol (C$_2$B$_{10}$H$_{12}$S, briefly will be 
called as carboranes or CT) derivatives are potential molecules for 
SAMs in various fields of applications. Due to their spherical 
stable nature, the dipole moment vectors and orientations can be easily 
modified by changing the position of the carbon atoms in 
the cage structure. This kind of material design gives the possibility to 
modulate the work function of the surface on which it is coated.\cite{Base2005, 
Base2008, Hohman2009,Hohman2010, Balaz2007, Base2010, Scholz2011, Lubben2011, 
Base2012, Kim2014,Serino2017, Yavuz2017} L\"{u}bben~\textit{et al.} 
modulated the work function of Ag(111) with pure and mixed SAM structures 
formed using 1,2 and 9,2 type carboranedithiol molecules 
(1,2-(HS$_2$)2–1,2-C$_2$B$_{10}$H$_{10}$ and 
9,12-(HS$_2$)2–1,2-C$_2$B$_{10}$H$_{10}$) having opposite dipole orientations 
on the silver surface.\cite{Lubben2011} Weiss~\textit{et al.} studied 
Au(111) with SAMs composed of 1,7 M1 and M9 type carboranes 
(1-HS-1,7-C$_2$B$_{10}$H$_{11}$ and 9-HS-1,7-C$_2$B$_{10}$H$_{11}$). 
\cite{Kim2014} They were able to systematically modulate the Au(111) work 
function in the 0.8 eV range with different coating ratios. When they transfer 
their coatings to OFET design, mixed SAM structures produced better yields. 
Also, the carboranes can be systematically functionalized with various ligand 
groups with different electronic properties.\cite{Serino2017} In addition, 
functionalized carboranes alter the electronic structure of the surface on which 
they are attached. As a result, SAMs with functionalized carboranes gives 
opportunity to systematically change the electronic and the surface properties 
of materials. 

Functionalized CT-SAMs have recently received an increased 
attention.\cite{Ito2005,Thomas2015, Serino2017} Yamamoto~\textit{et al.} 
brough insights into surface chemical behavior of functionalized 
para-carboranes containing different azo-benzene derivatives on the 
alkyl-coated Au surface.\cite{Ito2005} Thomas~\textit{et al.} examined 
geometries of carboxyl-functionalized para-carboranes on Au(111) by both 
experimental techniques and computational modeling.\cite{Thomas2015} 
Weiss~\textit{et al.} investigated the adhesion properties of carborane isomers 
with various anchor groups on Ge surface.\cite{Serino2017} The COOH anchor group 
was found to be more suitable than the thiol group for carborane based SAM 
structures on Ge surfaces. In addition, the work function of germanium with 
carborane carboxylate monolayers was shown to be modifiable in a 0.4 eV band 
without affecting the surface properties of germanium.

Utilization of functional groups with different electronic characteristics, 
particularly, for carborane-SAMs enables researchers to advance in selective 
bio-applications. For example, in a recent study, Neirynck~\textit{et al.} 
showed that RGD-functionalized carboranes on a full monolayer b-cyclodextrin 
coated Au surface could detect C2C12 type cells.\cite{Neirynck2015} In 
addition, some recent experimental studies have reported that the amphiphiles 
behavior of functionalized carborane derivatives is effective on some cancer 
cells.\cite{xiong2016amphiphilic, xiong2017stable}  

A thorough understanding of the electronic and structural properties of 
carborane based SAMs with functional groups is of great importance for 
development of novel applications. Computational studies are needed to bring 
new insights regarding the functionalized carborane-SAMs. In this study, we 
investigated the electronic and morphological structures of Au(111) in the 
presence of carborane isomers which are functionalized by COOH (as an acceptor 
group) and NH$_2$ (as a donor group) in the framework of dispersive corrected 
density functional theory (vdW-DFT) simulations. We examined the 
role of the functional groups on the molecular dipole moments and, therefore, 
on the adsorption characteristics of charboranethiols on Au(111). 
Steric requirements due to position of the functional groups on the 
carboranethiol isomers play an active role on hexagonal molecular 
arrangement on the gold surface. In addition, the influence of the functional 
groups with counteracting electron affinities on the molecular dipole moments 
is determined in relation to tunability the work function of gold in the cases 
of isolated and monolayer carboranethiols. 

\section{Computational Method}

Total energy density functional theory (DFT) calculations were 
performed within the framework of projector-augmented wave (PAW) 
method\cite{Joubert,Blochl} as implemented in Vienna Ab initio Simulation 
Package (VASP).\cite{Kresse, Furth} Single particle Kohn-Sham (KS) orbitals 
were expanded in the plane-wave basis up to a kinetic energy cut-off value of 
400 eV. The exchange-correlation (XC) many-body effects and the van der Waals 
interactions were included using the meta-GGA SCAN+rVV10 
functional.\cite{Peng2016}
 
In order to represent the Au surface, four layer slab models with (111) 
surface termination was constructed by cleaving from bulk gold. Computational 
supercells include the gold slab, carboranethiol adsorbate(s) on the top layer 
and a 12 {\AA} thick vacuum space along [111] direction. Each pure and 
functionalized carboranethiol isomer was considered as a single adsorbate on the
Au(111)-(5$\times$5) surface cell. The separation between the periodic images 
of a molecule is 13.72 {\AA} which allows us to consider the carboranes to be 
isolated on the (5$\times$5) structure. Possible adsorption sites were 
systematically examined for pure carborane derivatives 
previously.\cite{Mete2016} In the cases of functionalized CTs, atop, bridge and 
hollow sites were considered as the initial adsorption configurations. For the 
full monolayer, two possible conformations were taken into account. The first 
one involves a single CT on the (3$\times$3). This dense packing arrangement 
allows hydrogen bonding since the functional group of one molecule gets closer 
to the boron vertex on the adjacent molecule. In the second one, two CTs are 
placed such that the functional groups form dimers which is commensurate with 
the (6$\times$3) computational supercell. This conformation accepts a potential 
hydrogen bonding between the functional groups (COOH..COOH or NH$_2$..NH$_2$).  
 
Brillouin zone integrations were carried out using $\Gamma$-centered  
8$\times$8$\times$1, 4$\times$8$\times$1 and 5$\times$5$\times$1 $k$-point 
samplings for (3$\times$3), (6$\times$3) ve (5$\times$5) surface cells, 
respectively. A smearing parameter of 0.05 was adopted for Methfessel-Paxton 
(MP) scheme. Geometry optimizations of the model structures were obtained by 
minimization of the Hellmann-Feynman forces on each atom until a tolerance 
value of 10$^{-2}$ eV/{\AA} was reached in all spatial directions. The 
atoms at the bottom layer of the gold slab were frozen to their bulk positions. 

The dissociative chemisorption energies, E$_{\mathrm{c}}$, of pure and 
functionalized carboranethiol isomers on unreconstructed Au(111) with various 
periodicities can be calculated using,
\[
E_{\mathrm{c}}=\frac{E_{\mathrm{CT+Au(111)}}-E_{\mathrm{Au(111)}}-
n(E_{\mathrm{CT}}-E_{\mathrm{H}})}{n}
\]
\noindent where $E_{\mathrm{CT+Au(111)}}$ is the total energy of the supercell 
which contains the gold slab with $n$ number of CT adsorbates. 
$E_{\mathrm{Au(111)}}$ and $E_{\mathrm{CT}}$ are the total energy of the clean 
gold slab and the energy of a single CT molecule in a big box, respectively. 
The energy of an H atom, $E_{\mathrm{H}}=E_{\mathrm{H_2}}/2$, is calculated
from molecular hydrogen. 

\begin{figure}[htb]
\includegraphics[width=8.5cm]{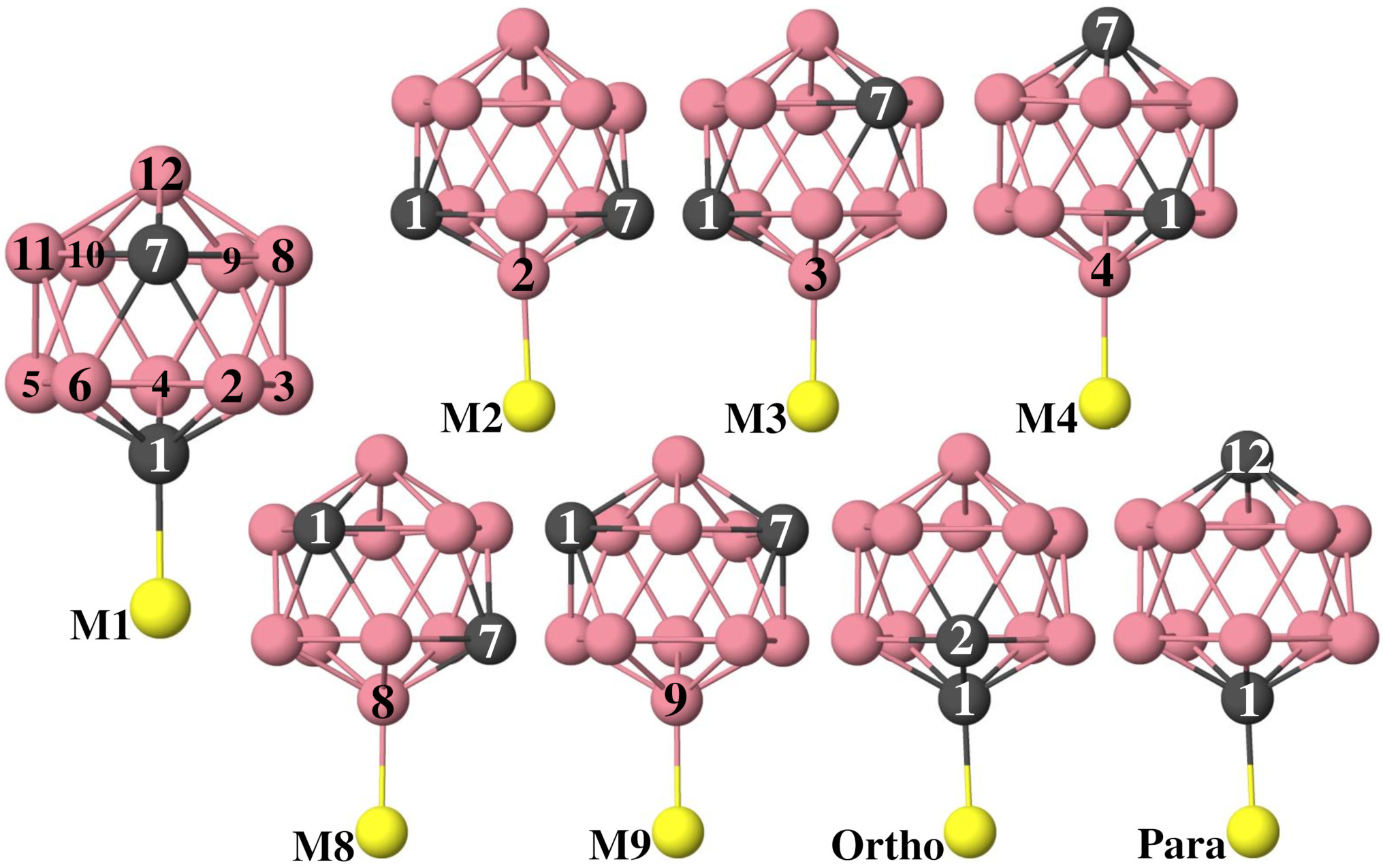}
\caption{Carboranethiol (CT) isomers are depicted in ball-and-stick 
form where pink, black and yellow balls represent boron, carbon and 
sulphur atoms. Hydrogen atoms are not shown for visual convenience. Labeling of 
meta isomers follows the attachment of thiol group on the carborane cage while 
keeping two carbon atoms at positions 1 and 7. The ortho (O) and para (P) 
variants differ from each other by the positioning of carbon atoms.}\label{fig1}
\end{figure}

The work function of Au(111) with functionalized CT adsorbates can 
be calculated as the difference between the vacuum level and the Fermi energy 
of the system. The vacuum level can be obtained from the planar-averaged 
electrostatic potential profiles along the surface normal 
($z$-direction), which is given as, 
\[
\overline{\rm V}(z)=\frac{1}{A}\,\int\hspace{-2mm}\int\limits_{\hspace{-3mm}\rm 
cell} \,\, V(x,y,z)\, dx\,dy  
\]
\noindent where $V(x,y,z)$ is the real space electrostatic 
potential and $A$ is the area of the corresponding surface cell.

\section{Results and Discussion}

Optimized geometries of carboranethiol isomers were previously obtained using 
DFT calculations.~\cite{Mete2016} We follow the same labeling scheme of those 
positional isomers as presented in Fig.~\ref{fig1}. In particular, 
meta-carboranethiols are labeled with respect to the cage atom to which thiol 
group is attached. Recent experiments showed that ordered monolayers of COOH 
functionalized M1 and M9 carboranethiols are commensurate with the (5$\times$5) 
and (3$\times$3) unit cells of unreconstructed Au(111).\cite{Goronzy} 

\begin{figure}[htb]
\includegraphics[width=8.5cm]{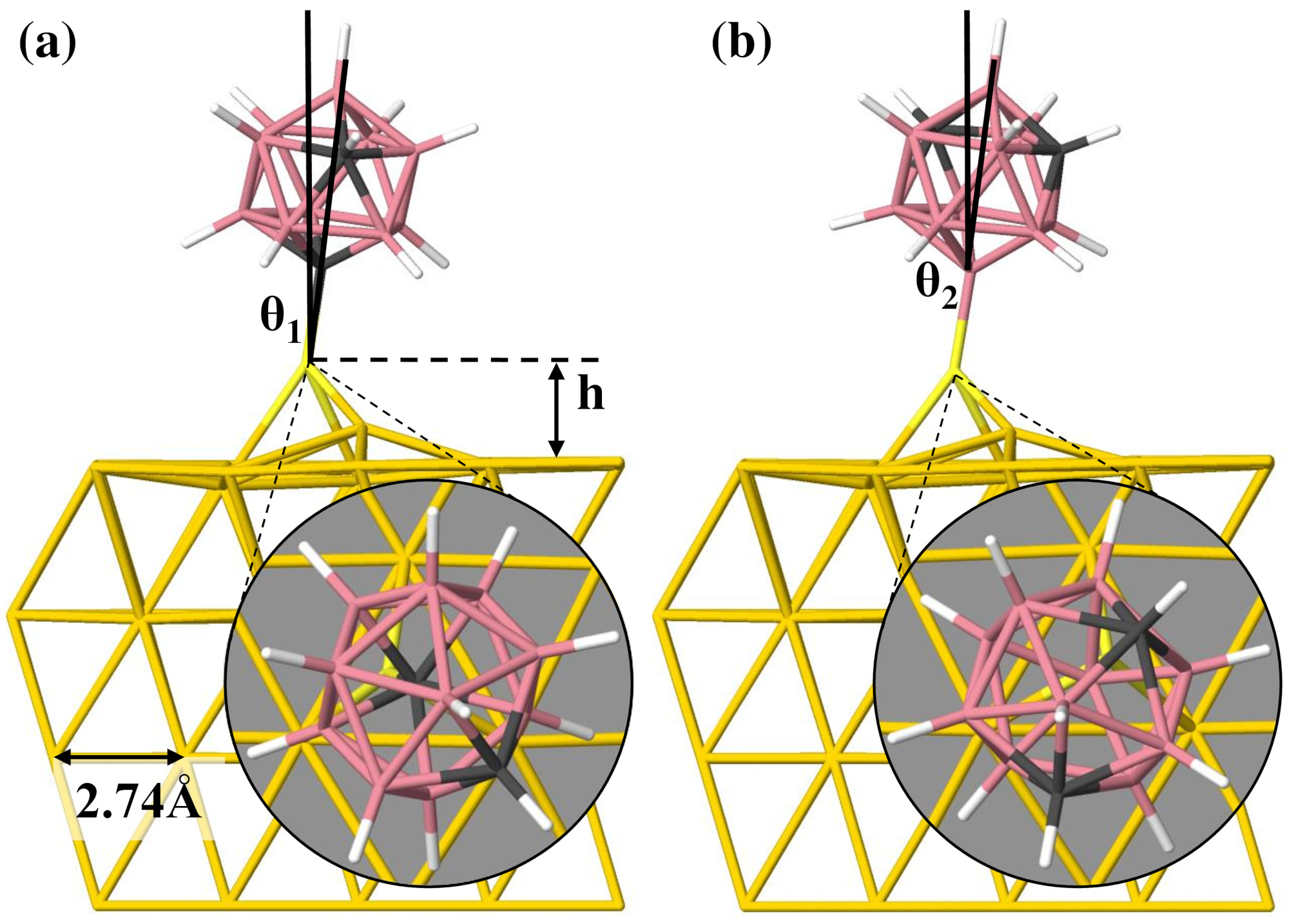}
\caption{Side and top (insets) views of (a) M1 and (b) M9 carboranethiol 
isomers on 5$\times$5-Au(111) surface cell optimized using the SCAN+rVV10 DFT 
functional.  The tilting angles of the S-cage bond and the major axis of the 
cage (with respect to the surface normal) are depicted on the models as 
$\theta_{1}$ and $\theta_{2}$, respectively. The height, $h$, indicates the 
separation between the S atom and the gold surface plane.}\label{fig2}
\end{figure}

\subsection{Isolated Carboranethiols on Au(111)}

\begin{figure*}[htb]
\includegraphics[width=16.5cm]{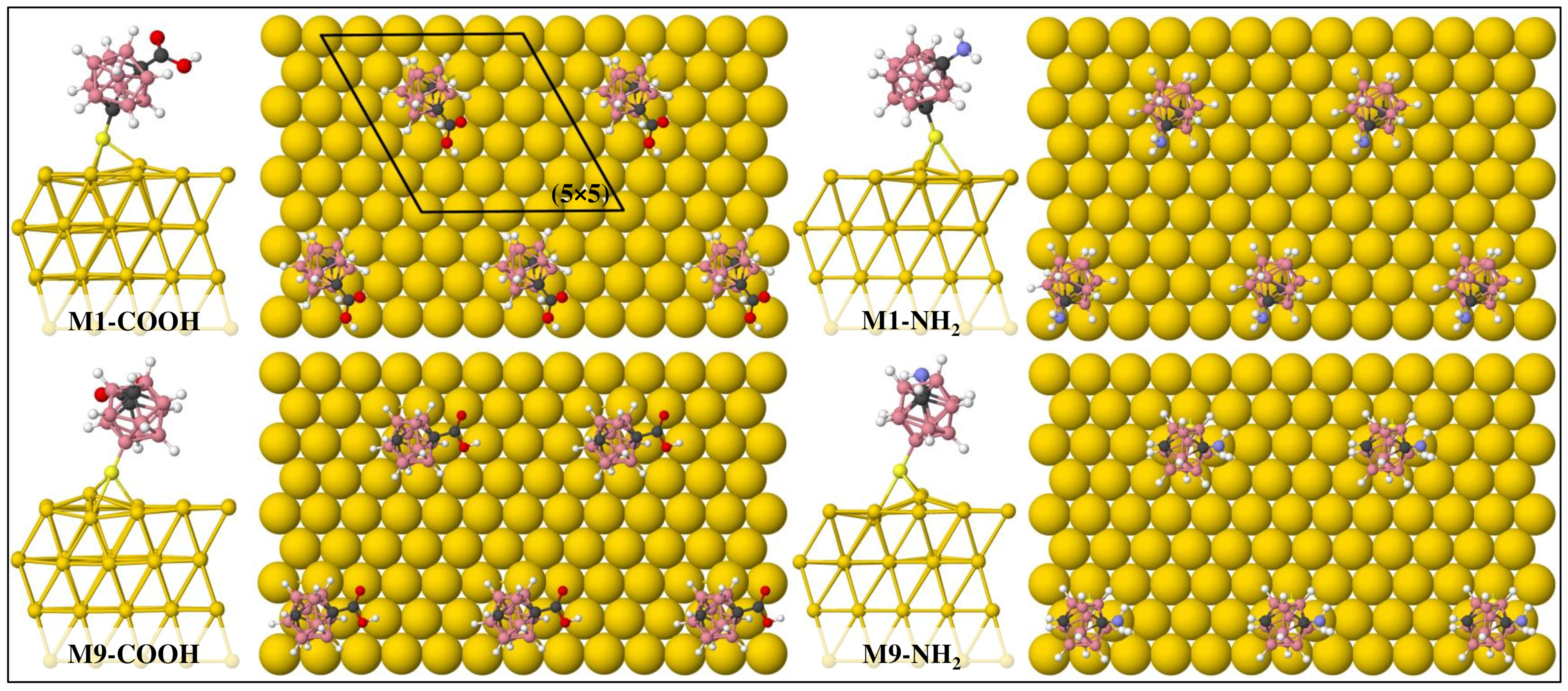}
\caption{The top and side views of COOH- and NH$_2$-functionalized M1 and M9 
carboranes with (5$\times$5) periodicity on Au(111). The structures were 
optimized using the SCAN+rVV10 functional. Yellow, gray, pink, red, and blue 
balls on the molecules represent S, C, B, O, and N atoms, respectively.}
\label{fig3}
\end{figure*}

A single isolated molecule was considered on Au(111)-(5$\times$5) structure. 
The chemisorption of pure, COOH- and NH$_2$-functionalized CT isomers 
were systematically examined by considering possible initial adsorption sites 
which involve bridge, top and hollow positions on the (111) 
surface of gold as described for alkanethiols previously.\cite{Mete2017} 
Geometry optimization computations based on the SCAN+rVV10 functional show that 
the thiol part of carborane derivatives plays a dominant role on the 
chemisorption site. The CT molecules are dragged by the S heads to a position 
on a triangle formed by three surface gold atoms, the hollow position. The S 
atom gets closer to bridge site forming two bonds with the bridging Au atoms. 
In this configuration, the S atom also forms a relatively shorter bond with the 
Au atom across the bridge site by lifting it up from the surface plane as shown 
in Fig.~\ref{fig2}.  

\begin{table}[htb]
\caption{Structural parameters and adsorption energies of pure and 
functionalized carboranethiol (CT) isomers on Au(111)-(5$\times$5) surface cell 
optimized by DFT calculations using the SCAN+rVV10 functional. Adsorption 
energies ($E_{\mathrm{c}}$), bond lengths and heights
($d_{\mathrm{S}\textrm{-}\mathrm{Au}}$, 
$d_{\mathrm{S}\textrm{-}\mathrm{Cage}}$, h), and tilting angles ($\theta_{1}$, 
$\theta_{2}$) are given in units of eV, {\AA}, and degrees, respectively.}
\label{table1}\vskip1mm
\begin{tabular}{c|c|c|c|c|c|c} \hline
\hspace{2mm}\mbox{}Molecule\hspace{2mm}\mbox{}&
\hspace{2mm}\mbox{}$E_{\mathrm{c}}$\hspace{2mm}\mbox{} &
$d_{\mathrm{S}\textrm{-}\mathrm{Cage}}$&
\hspace{7mm}\mbox{}$d_{\mathrm{S}\textrm{-}\mathrm{Au}}$\hspace{7mm}\mbox{}& 
\hspace{2mm}\mbox{}$\theta_{1}$\hspace{2mm}\mbox{} & 
\hspace{2mm}\mbox{}$\theta_{2}$\hspace{2mm}\mbox{} & 
\hspace{2mm}\mbox{}h\hspace{2mm}\mbox{} \\ \hline
M1 &-0.93~& 1.81 &2.32, 2.65, 2.71& 12.2 & ~8.2 & 1.86\\			
M2 &-0.86 & 1.85 & 2.33, 2.62, 2.66 & ~9.3 & ~8.0 & 1.73\\
M3 &-0.97 & 1.87 & 2.33, 2.62, 2.63 & 11.8 & 13.4 & 1.74\\
M4 &-1.03 & 1.87 & 2.33, 2.48, 2.63 & 12.4 & ~9.5 & 1.63\\
M8 &-0.98 & 1.87 & 2.32, 2.52, 2.77 & 13.3 & ~7.9 & 1.75\\
M9 &-1.03 & 1.88 & 2.33, 2.51, 2.60 & 12.2 & 11.1 & 1.71\\
O  &-0.95 & 1.81 & 2.36, 2.48, 2.85 & 25.2 & 16.6 & 1.91\\
P  &-0.88 & 1.82 & 2.32, 2.75, 2.82 & 12.3 & ~7.6 & 2.02\\ \hline
M1-COOH &-1.20 &1.81 &2.36, 2.43, 2.92 &29.6 &21.0  &1.91 \\ 
M2-COOH &-1.02 &1.85 &2.34, 2.47, 2.91 &22.5 &14.4  &1.79 \\ 
M3-COOH &-1.60 &1.86 &2.42, 2.42, 3.05 &42.3 &38.0  &1.96 \\
M4-COOH &-1.31 &1.86 &2.33, 2.42, 2.92 &33.0 &19.8  &1.86 \\ 
M8-COOH &-1.06 &1.86 &2.38, 2.51, 2.82 &26.8 &15.2  &1.92 \\
M9-COOH &-1.37 &1.87 &2.34, 2.42, 2.90 &30.6 &21.8  &1.84 \\
~~O-COOH &-1.09 &1.80 &2.33, 2.55, 2.97 &19.7 &15.0  &1.93 \\
~~P-COOH &-1.29 &1.81 &2.33, 2.43, 2.96 &32.8 &32.3  &1.93 \\ \hline
M1-NH$_2$ &-1.02 &1.81 &2.34, 2.85, 2.91 &28.4 &22.1 &2.12\\
M2-NH$_2$ &-0.78 &1.84 &2.34, 2.50, 2.68 &14.0 &10.2 &1.69\\
M3-NH$_2$ &-1.21 &1.87 &2.35, 2.36, 2.86 &27.7 &25.2 &2.07\\
M4-NH$_2$ &-1.22 &1.86 &2.37, 2.41, 2.83 &33.1 &30.9 &1.71\\
M8-NH$_2$ &-1.14 &1.86 &2.33, 2.35, 2.81 &20.3 &~9.7 &1.92\\
M9-NH$_2$ &-1.20 &1.88 &2.33, 2.35, 2.91 &27.1 &20.5 &2.15\\
~~O-NH$_2$ &-1.00 &1.78 &2.33, 2.90, 3.09 &29.6 &28.6 &2.30\\
~~P-NH$_2$ &-1.06 &1.81 &2.36, 2.72, 2.88 &29.9 &23.8 &2.04\\ \hline
\end{tabular}
\end{table}

\begin{table}[htb]
\caption{Dipole moment components (in units of Debye) of carboranethiol (CT) 
molecules, as adsorbed on the Au(111) with the (5$\times$5) periodicity, 
calculated using the SCAN+rVV10 functional. The $z$-axis aligns with [111] 
direction and the $x$-axis is oriented along [101] direction. Work function 
values ($\Phi$ in eV) calculated for Au(111) with CT adsorbates (pure and 
functionalized) in the (5$\times$5) surface structure. Charge transfer values 
($\Delta Q$ in $e$) from the gold surface to individual molecules are given for 
their corresponding chemisorption geometries.}
\label{table2}\vskip1mm
\begin{tabular}{c|c|c|c|c|c} \hline
\hspace{3mm}\mbox{}Molecule\hspace{3mm}\mbox{}&
\hspace{4mm}\mbox{}$\mu_{x}$\hspace{5mm}\mbox{}&
\hspace{4mm}\mbox{}$\mu_{y}$\hspace{5mm}\mbox{}&
\hspace{4mm}\mbox{}$\mu_{z}$\hspace{5mm}\mbox{}&
\hspace{3mm}\mbox{}$\Phi$\hspace{3mm}\mbox{}&
\hspace{3mm}\mbox{}$\Delta Q$\hspace{3mm}\mbox{}\\ \hline
M1 &-2.187 &-0.398 &~0.124 & 5.67 & 0.117\\			
M2 &-0.943 &-0.938 &-0.345 & 5.66 & 0.086\\
M3 &~2.249 &-0.584 &~2.121 & 5.54 & 0.060\\
M4 &-2.291 &~0.557 &~3.176 & 5.49 & 0.018\\
M8 &-0.545 &~2.205 &~2.093 & 5.53 & 0.047\\
M9 &~0.694 &-1.189 &~4.389 & 5.40 & 0.005\\
O  &-1.680 &~1.231 &-1.567 & 5.77 & 0.130\\
P  &-0.176 &-0.155 &~1.670 & 5.65 & 0.149\\ \hline
M1-COOH &~0.549 &-2.207 &-0.721 & 5.74 & 0.091 \\
M2-COOH &~0.812 &-2.617 &-0.196 & 5.72 & 0.107 \\
M3-COOH &-1.252 &-2.413 &~1.195 & 5.70 & 0.042 \\
M4-COOH &~0.207 &~1.870 &~3.517 & 5.59 & 0.021 \\
M8-COOH &-1.823 &-1.201 &~0.825 & 5.69 & 0.086 \\
M9-COOH &~0.823 &-1.605 &~4.459 & 5.49 &\hspace{-1mm}-0.007 \\
~~O-COOH &-0.917 &-1.939 &-1.959 & 5.80 & 0.200 \\
~~P-COOH &~1.325 &~0.522 &~1.227 & 5.71 & 0.083 \\ \hline
M1-NH$_2$ &-1.686 &-2.133 &~1.038 & 5.80 & 0.119\\
M2-NH$_2$ &~1.033 &-1.611 &~0.202 & 5.79 & 0.049\\
M3-NH$_2$ &~2.969 &-1.649 &~2.034 & 5.69 &\hspace{-1mm}-0.010\\
M4-NH$_2$ &-4.780 &-0.113 &~2.108 & 5.63 & 0.012\\
M8-NH$_2$ &~1.266 &~1.189 &~3.580 & 5.57 & 0.014\\
M9-NH$_2$ &~0.464 &-0.201 &~4.121 & 5.54 &\hspace{-1mm}-0.029\\
~~O-NH$_2$ &-2.518 &-0.574 &-2.735 & 5.92 & 0.115\\
~~P-NH$_2$ &-0.425 &-0.746 &~3.200 & 5.75 & 0.109\\ \hline
\end{tabular}
\end{table}

The (5$\times$5) structure makes enough room for a separation of 13.72 {\AA} 
between the periodic images of the S atoms, which corresponds to a low coverage 
adsorption case as depicted in Fig.~\ref{fig3}. The S-Au bond lengths are 
consistent with covalent bond distances as given in Table~\ref{table1}. Upon 
adsorption, the bond between the S atom and the carborane cage becomes tilted 
with respect to the surface normal, which is referred as $\theta_{1}$ as 
shown in Fig.~\ref{fig2}. In most cases, the tilting of the cage itself, 
$\theta_{2}$, differs from $\theta_{1}$ as seen in Table~\ref{table1}. 

The bond length between the S atom and the carborane cage is affected by the 
attachment of the carboxyl and amine groups, only slightly. The bond 
distances are noticeably smaller for M1, Ortho (O) and Para (P) derivatives 
where S forms the bond with C atom of the cage. Therefore, the atomic species
being either boron or carbon, to which the thiol group is attached on the 
carborane cage, can be identified as the main factor on the S-cage 
bonding. 

On the other hand, both the functional groups and the molecular dipole 
moments (depending on the isomer type) become more important on the molecular 
adsorption and tilting angles on the (111) surface of gold. For instance, 
functionalized CT isomers get relatively more tilted on the gold surface at 
both low and high coverages. This situation is related to the molecular 
symmetry. The cage geometry of pure CT isomers are essentially the same. 
However, the terminal groups break the cylindrical symmetry. Therefore, 
functional groups not only bring additional charge but also lead to an 
asymmetrical charge distribution over the molecules. In addition, as the 
tilting increases, the carborane cage which has an outer diameter of $\sim$5.3 
{\AA} comes closer to the surface. As a results, a long range electronic 
interaction between the cage and the surface atoms contributes to the 
adsorption energy. 

Four main factors can be considered, which influence the adsorption 
characteristics of the isolated CT molecules. These are the bonding between the 
thiol and surface gold atoms, the molecular dipole moment, the functional 
group, and the van der Waals interaction between the cage and the gold surface. 
The adsorption energies presented in Table~\ref{table1} reveal that the 
strength and covalency of S-Au bonds are affected by the functional groups. 
For instance, the binding of M3-COOH variant is significantly stronger on the 
(5$\times$5) structure. This case is also a good example which reflects the 
contribution of van der Waals interaction between the cage and the metal 
surface on its adsorption energy. The tilting of M3-COOH on the (5$\times$5) 
structure is significantly larger among the other derivatives such that the 
closest distance from the cage to the gold surface becomes as small as 
2.58 {\AA}.   

When the tilt angles, $\theta_{1}$ and $\theta_{2}$, of carborane derivatives 
in their computationally optimized structures are compared, CT molecules with 
amine or carboxyl functional groups appear to be tilted more relative to their 
pure counterparts. In addition, molecules which get strongly adsorbed on the 
metal surface have relatively higher tilting angles. Hence, presence of the 
functional groups can be related to the angle of tilting which shows a 
correlation with the binding energy, or with the stability, of CT derivatives 
on the metal surface. Among the pure isomers, the only exception seems to 
be the ortho (O) case. The nearest neighbor placement of C atoms at the 
positions 1 and 2 as shown in Fig.~\ref{fig1} causes an increased local 
electron density which leads to a molecular dipole moment, with a negative $z$ 
component (in Table~\ref{table2} and Fig.~\ref{fig3}), pointing toward the gold 
surface. This leads to a higher tilting angle relative to other pure isomers. 

The position of functional groups change with the isomers depending on the 
positions of C atoms on the cage. In most cases, these groups are exposed on 
the surface. On the other hand, in some cases, these groups get closer to the 
metal surface upon chemisorption. For M2-COOH and O-COOH, the distance from 
the carboxyl to the surface is 4.05 {\AA} and 4.09 {\AA}, respectively. For 
M2-NH$_2$ and O-NH$_2$, these distances become 4.50 {\AA} and 4.06 {\AA}, 
respectively. Although these values are still large to consider a strong and 
direct interaction between the functional groups and the gold surface, 
positioning of the carboxyl and amine groups have an important role on the 
adsorption characteristics. Not only the tilting angles but also 
the functional groups increase steric demands of the CTs in the SAMs.

\begin{figure}[htb]
\includegraphics[width=8.5cm]{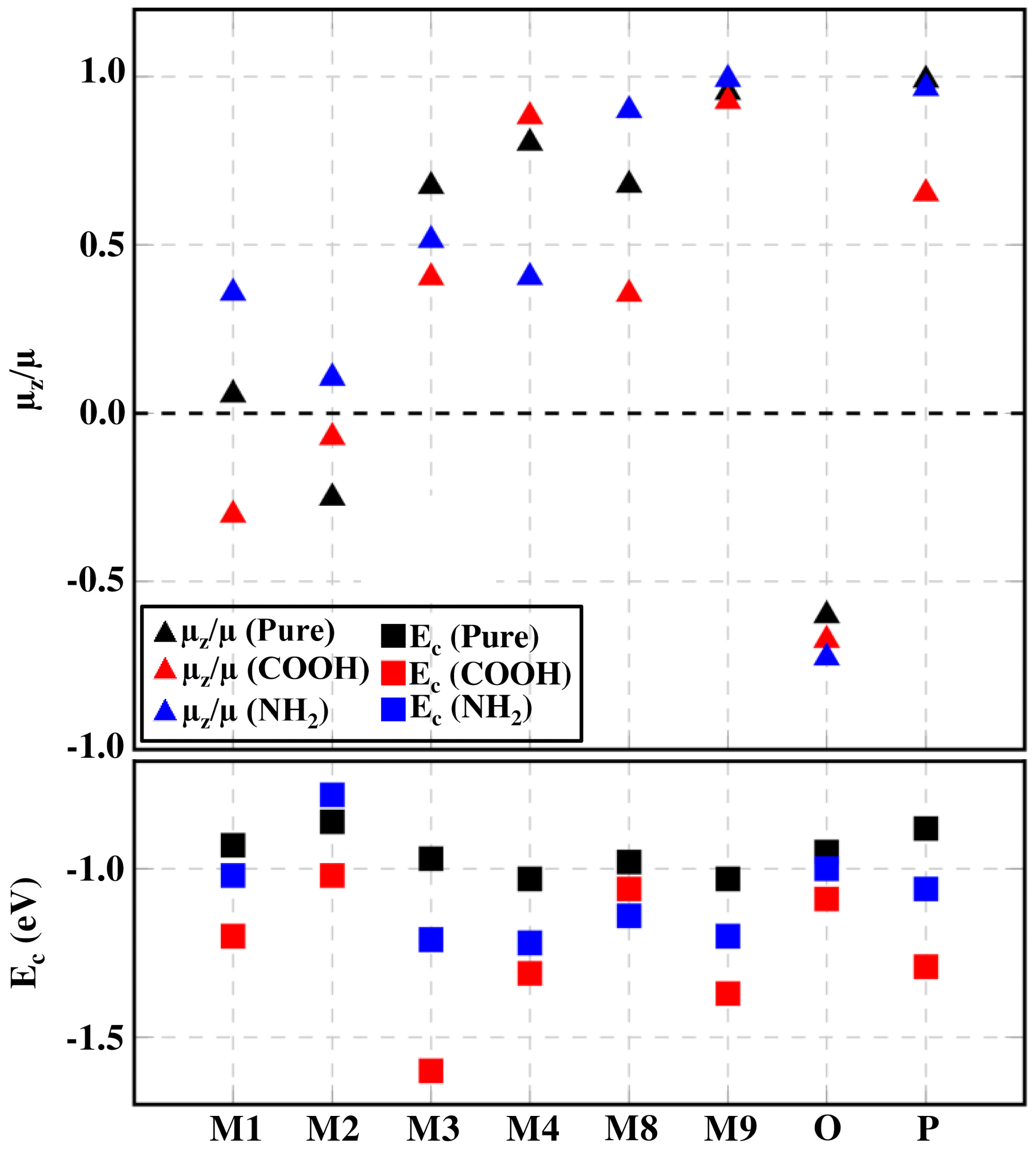}
\caption{DFT-calculated $z$-components (corresponding to [111] direction gold 
surface) of molecular dipoles normalized to their net moments ($\mu_z/\mu$) and 
the chemisorption energies ($E_{\mathrm{c}}$) of pure and functionalized 
carboranethiols on the Au(111)-(5$\times$5) using the SCAN+rVV10 
functional.}\label{fig4}
\end{figure}

Molecular dipole moments were calculated in order to explore their effect on 
the binding energies. For this purpose, CT molecules were considered in big 
computational boxes for which the $z$- and $x$-axes are chosen to coincide with
the [111] and [101] directions of the gold slab. Then, the alignment of each CT 
molecule was kept as it was adsorbed on the gold surface. The results for 
(5$\times$5) cases are presented in Table~\ref{table2}. Previous studies 
emphasized the role of molecular dipoles on the formation and stability of CTs 
on Au(111) by considering pure M1 and M9 positional isomers. Although they have 
similar geometrical shapes, M1 and M9 possess molecular dipole orientations 
almost parallel and perpendicular to the gold surface with moments of 1.06 D 
and 4.08 D in the gas phase, respectively.\cite{Hohman2009} Our results show 
that M1 and M9 isomers essentially keep their gas phase dipole orientations 
even though they get sligthly tilted after chemisorption on the gold surface at 
low density (5$\times$5) hexagonal packing. The binding energies show a 
correlation with both the dipole moments and their orientations. A larger 
dipole with a sizable component along surface normal leads to a relatively 
stronger chemisorption. The dipole components perpendicular to the surface 
normalized to the net dipole moments ($\mu_z/\mu$) and chemisorption energies 
($E_{\mathrm{c}}$) are depicted in Fig.~\ref{fig4} for all adsorption cases 
of pure isomers on the (5$\times$5) supercell. Indeed, the weakest adsorption 
occurs for M2 isomer among the pure and functionalized cases, which has the 
smallest dipole component along [111]. As a trend for the cases with moderate 
tilting angles, a molecular dipole parallel to the metal surface gives a weaker 
binding relative to that perpendicular to the surface in the +$z$-direction 
([111] direction relative to underlying gold). 

The CT molecules bind significantly stronger to the gold surface 
with the attachment of both of the functional groups. In fact, the nominal 
charge states of COOH and NH$_2$ groups in a chemical environment are 
different, the former being an electron acceptor and the latter being an 
electron donor group. The amine moiety enhances adsorption of carboranes 
isomers on gold relative to their corresponding pure cases with the only 
exception of M2. In particular, carboxylic acid group makes carboranes to bind 
to the gold surface even slightly stronger relative to the amine group. The 
change in the binding energy is in the range between 0.14 eV (for O) and 0.63 
eV (for M3). The role of charge transfers from CT molecules to the gold 
substrate were considered and the calculated values are given in 
Table~\ref{table2}. It appears that CT molecules, functionalized or not, get 
slightly oxidized upon adsorption in most of the cases. The amount of charge 
transfer from the gold surface to the CTs seems to be affected mainly by the 
positional isomerization. The results indicate that the type of the terminal 
group is less effective in S-Au bonding characteristics.

DFT calculations using SCAN+rVV10 functional predicts a work function of the 
bare Au(111) as 5.41 eV\cite{Patra} which is reasonably acceptable when 
compared with the experimentally estimated value of 5.33$\pm$0.06 
eV.\cite{Derry}  Experimental studies reported a change in the work function 
with M9 monolayers on gold.\cite{Hohman2009,Kim2014} The influence was 
considerably smaller in the case of M1.\cite{Hohman2009} The change in the work 
function is attributed to the molecular dipole moments and their orientations 
on the gold surface. Since M9 has a considerably larger dipole moment relative 
to that of M1, and since its dipole points almost perpendicular to the surface, 
M9 adsorption has a larger influence on the work function. In fact, adsorbates 
with a dipole with a positive pole pointing away from the surface cause a 
decrease in the work function. The effect, even, reverses if the negative pole 
points away from the surface.\cite{Crispin,Zehner,Ishii} In order to examine 
the role of the functional groups on the work function, we first considered all 
the isomers on the (5$\times$5) structure as given in Table~\ref{table2}. In 
consistency with the observations of Kim~\textit{et al.}\cite{Kim2014}, the 
lowest work function was obtained for M9 among pure CT isomers. Our 
calculations indicate an increase in the work function by chemisorption of CT 
molecules with dipoles aligning parallel to the gold surface as in the cases of 
M1 and M2. The most dramatic increase happens for the adsorption of ortho-CT 
which has a dipole almost opposite to M9. A similar influence of dipole moments 
on the work function is also seen for the functionalized CTs. The both 
functional groups lead to an increase in the work function of gold relative to 
the pure cases. The change is noticeably larger with the amine functional group 
relative to carboxyl functional group. Adsorption of M4-COOH and M9-COOH gives 
considerably lower work function among the carboxyl functionalized CTs. 
Similarly, M8-NH$_2$ and M9-NH$_2$ isomers have the two lowest the work 
functions among the amine functionalized CTs as they possess the largest dipole 
moment component along surface normal. The results for the isolated cases 
indicate that pure and functionalized derivatives of M1, M2 and O CTs cause 
noticeably higher work function among the other isomers, as expected, depending 
on their dipole moments and orientations. The functional groups have an 
observable effect which allows a wider range for the tunability of the work 
function of gold.

\subsection{Carboranethiol Monolayers on Au(111)}

Densely packed monolayers of carboranethiols are commensurate with the 
(3$\times$3) unit cell of Au(111) where nearest-neighbor separation is 8.23 
{\AA}. Due to shorter spacing of molecules and steric requirements in the 
presence of functional groups, tilting angles can be expected to be smaller 
than that of the isolated cases. The M1 and M9 isomers were taken into account 
for SAM structures by many experiments in the first place due to their dipoles 
which are oriented parallel and perpendicular to the gold surface, 
respectively. Therefore, we focus on these isomers which reflect contrasting 
dipole orientations as favorable candidates to investigate the effect of 
carboxyl and amine functional groups on the characteristics of SAMs. Moreover, 
the adsorption geometries and binding energies can be compared in relation 
to possible dipole-dipole interactions in the monolayers. 

\begin{figure*}[t]
\includegraphics[width=16.5cm]{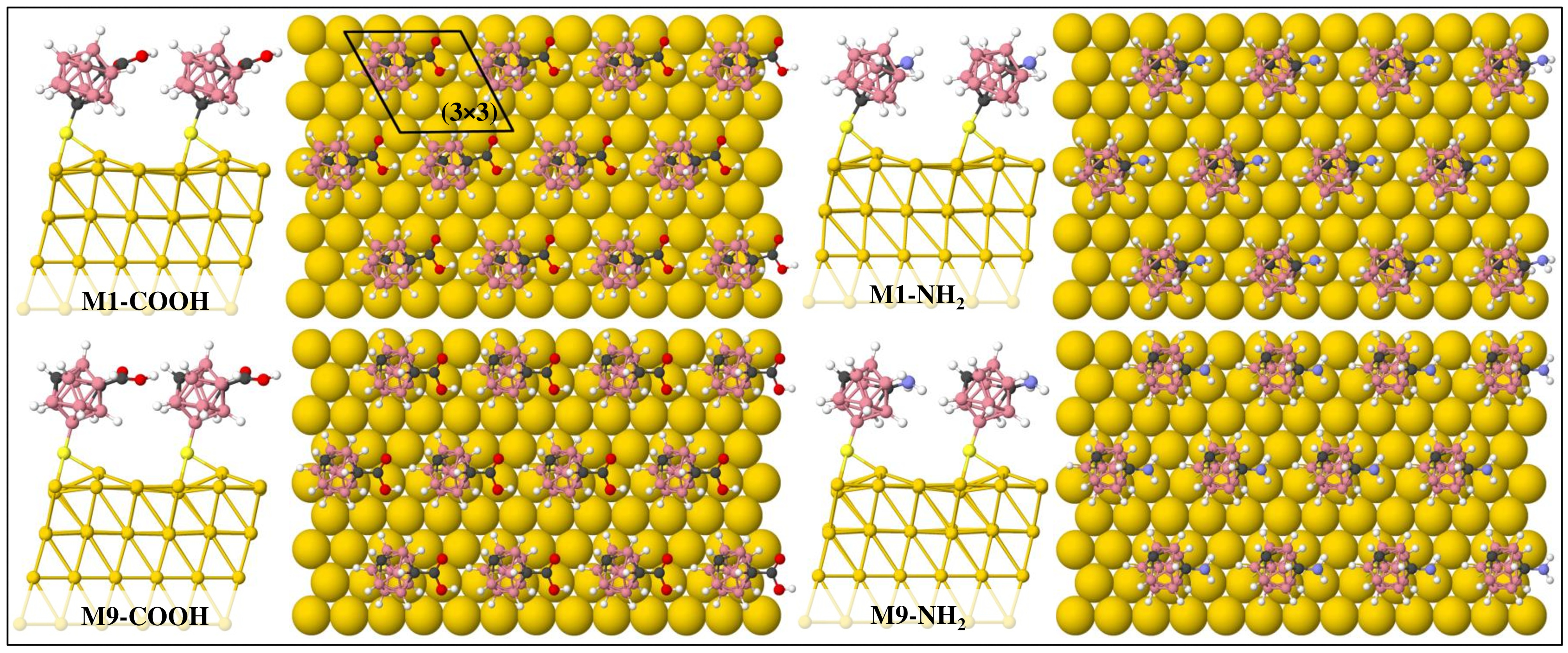}
\caption{The top and side views of COOH- and NH$_2$-functionalized M1 and M9 
carboranes with (3$\times$3) periodicity on Au(111). The structures were 
optimized using the SCAN+rVV10 functional. Yellow, gray, pink, red, and blue 
balls on the molecules represent S, C, B, O, and N atoms, respectively.}
\label{fig5}
\end{figure*}

On the (3$\times$3) structure two possible initial configurations can be 
considered. In the first one molecules are tilted backwards to expose the 
functional groups on the monolayer to the environment. The second one 
corresponds to the molecule tilting forward as shown in Fig.~\ref{fig5}. In 
this conformation each CT molecule with the functional group lean towards the 
neighboring molecule allowing a possibility of hydrogen interaction in the 
lateral direction. For M1-COOH, the latter configuration is energetically 0.1 
eV more favorable than the former one. Hence, computationally both structures 
are expected to be stable. Experiments reported that the two conformations are 
indistinguishable under STM measurements.\cite{Goronzy} An initial 
configuration with the first conformation relaxes to the second one in the 
cases of M9-COOH, M1-NH$_2$ and M9-NH$_2$ on the (3$\times$3) unit cell.

\begin{figure*}[t]
\includegraphics[width=16.5cm]{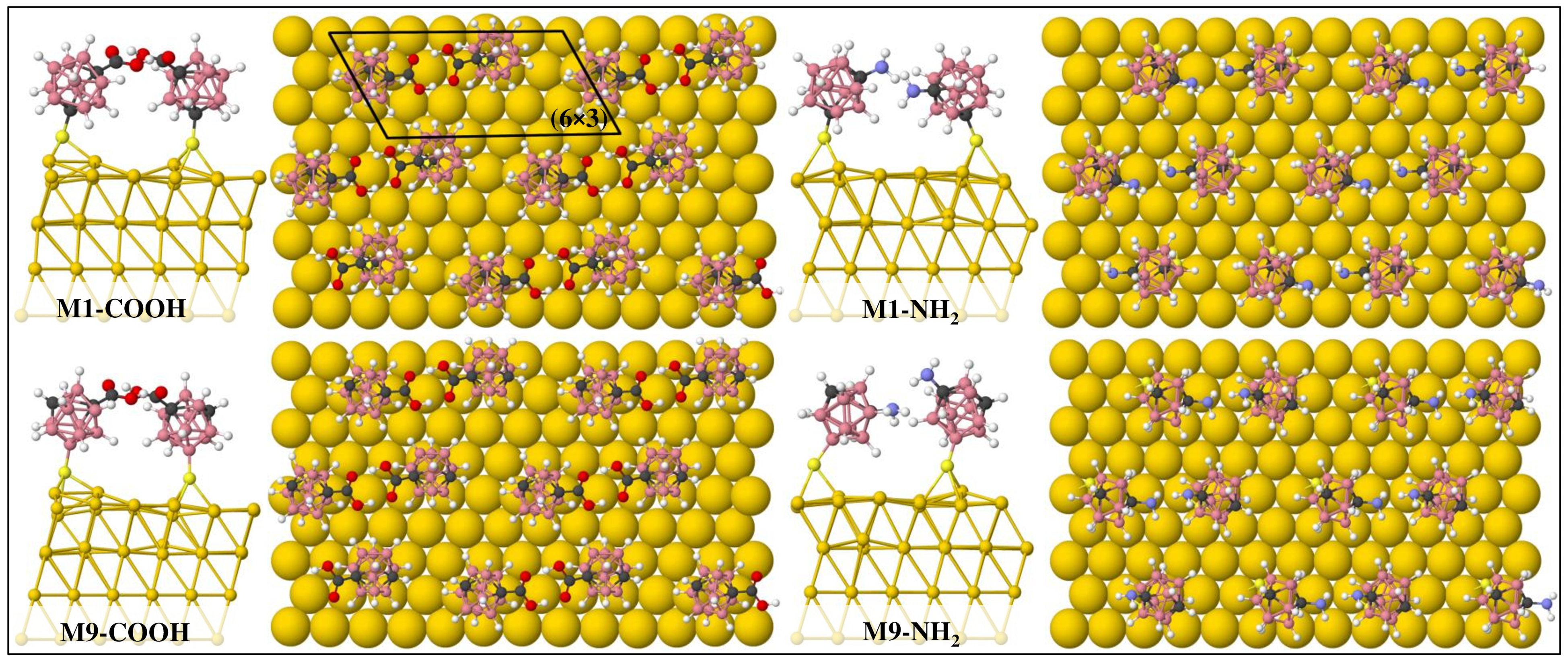}
\caption{The top and side views of COOH- and NH$_2$-functionalized M1 and M9 
carboranes with (6$\times$3) periodicity on Au(111). The structures were 
optimized using the SCAN+rVV10 functional. Yellow, gray, pink, red, and blue 
balls on the molecules represent S, C, B, O, and N atoms, respectively.}
\label{fig6}
\end{figure*}

\begin{table}[htb]
\caption{Structural parameters and energetics of pure and functionalized M1 and 
M9 carboranes on the (3$\times$3) and (6$\times$3) surface cells with respect 
to Au(111) calculated using the SCAN+rVV10 XC functional. Adsorption 
energies ($E_{\mathrm{c}}$), bond lengths and heights
($d_{\mathrm{S}\textrm{-}\mathrm{Au}}$, 
$d_{\mathrm{S}\textrm{-}\mathrm{Cage}}$, h), and tilting angles ($\theta_{1}$, 
$\theta_{2}$) are given in units of eV, {\AA}, and degrees, respectively.} 
\label{table3}\vskip1mm
\begin{tabular}{c|c|c|c|c|c|c} \hline
\hspace{2mm}\mbox{}Molecule\hspace{2mm}\mbox{}&
\hspace{2mm}\mbox{}$E_{\mathrm{c}}$\hspace{2mm}\mbox{} &
$d_{\mathrm{S}\textrm{-}\mathrm{Cage}}$&
\hspace{7mm}\mbox{}$d_{\mathrm{S}\textrm{-}\mathrm{Au}}$\hspace{7mm}\mbox{}& 
\hspace{2mm}\mbox{}$\theta_{1}$\hspace{2mm}\mbox{} & 
\hspace{2mm}\mbox{}$\theta_{2}$\hspace{2mm}\mbox{} & 
\hspace{2mm}\mbox{}h\hspace{2mm}\mbox{} \\ \hline
\multicolumn{7}{c}{CT/Au(111)-(3$\times$3)} \\ \hline
M1 &-1.31 &1.80 &2.31, 2.66, 2.75 &10.7 &~5.2  &1.97\\
M9 &-1.23 &1.85 &2.31, 2.62, 2.63 &~7.1  &~2.1  &1.91\\ \hline
M1-COOH  &-1.57 &1.80 &2.33, 2.73, 2.85 &21.6 &16.7 &2.17\\
M9-COOH  &-1.60 &1.86 &2.33, 2.59, 2.76 &15.5 &10.5 &2.09\\ \hline
M1-NH$_2$  &-1.51 &1.80 &2.34, 2.78, 2.92 &30.1 &24.9 &2.25\\
M9-NH$_2$  &-1.45 &1.86 &2.34, 2.53, 2.87 &23.1 &15.4 &2.14\\ \hline
\multicolumn{7}{c}{CT/Au(111)-(6$\times$3)} \\ \hline
M1 &-1.38 &
\begin{tabular}{c} 1.81 \\ 1.80 \end{tabular}&
\begin{tabular}{c} 2.34, 2.95, 2.97\\2.32, 2.97, 3.17\end{tabular}&
\begin{tabular}{c} 28.3 \\ 29.6 \end{tabular}&
\begin{tabular}{c} 22.8 \\ 24.1 \end{tabular}&
\begin{tabular}{c} 2.36 \\ 2.42 \end{tabular}\\[4mm] 
M9 &-1.38 &
\begin{tabular}{c} 1.81 \\ 1.81 \end{tabular}&
\begin{tabular}{c} 2.37, 2.75, 2.82\\2.40, 2.65, 2.84\end{tabular}&
\begin{tabular}{c} 30.3 \\ 30.5 \end{tabular}&
\begin{tabular}{c} 27.3 \\ 25.4 \end{tabular}&
\begin{tabular}{c} 2.26 \\ 2.15 \end{tabular}\\ \hline
M1-COOH &-1.91 &
\begin{tabular}{c} 1.81 \\ 1.81 \end{tabular}&
\begin{tabular}{c} 2.33, 2.36, 2.98\\2.33, 2.53, 2.80\end{tabular}&
\begin{tabular}{c} 30.1 \\ 17.3 \end{tabular}&
\begin{tabular}{c} 23.5 \\ ~8.7 \end{tabular}&
\begin{tabular}{c} 2.36 \\ 1.97 \end{tabular}\\[4mm]
M9-COOH &-1.90 &
\begin{tabular}{c} 1.86 \\ 1.86 \end{tabular}&
\begin{tabular}{c} 2.33, 2.36, 2.86\\2.39, 2.52, 2.80\end{tabular}&
\begin{tabular}{c} 20.5 \\ 21.2 \end{tabular}&
\begin{tabular}{c} 12.4 \\ 13.9 \end{tabular}&
\begin{tabular}{c} 2.24 \\ 1.87 \end{tabular}\\ \hline
M1-NH$_2$ &-1.59 &
\begin{tabular}{c} 1.80 \\ 1.80 \end{tabular}&
\begin{tabular}{c} 2.36, 2.91, 2.94\\2.35, 2.93, 3.00\end{tabular}&
\begin{tabular}{c} 32.2 \\ 33.6 \end{tabular}&
\begin{tabular}{c} 27.2 \\ 28.4 \end{tabular}&
\begin{tabular}{c} 1.87 \\ 2.02 \end{tabular}\\[4mm]
M9-NH$_2$ &-1.55 &
\begin{tabular}{c} 1.86 \\ 1.86 \end{tabular}&
\begin{tabular}{c} 2.37, 2.88, 2.98\\2.35, 2.74, 2.86\end{tabular}&
\begin{tabular}{c} 36.3 \\ 28.4 \end{tabular}&
\begin{tabular}{c} 28.8 \\ 20.5 \end{tabular}&
\begin{tabular}{c} 2.29 \\ 2.21 \end{tabular}\\ \hline
\end{tabular}
\end{table}

\begin{table}[htb]
\caption{Dipole moment components ($\mu_x$,$\mu_y$,$\mu_z$ in D) 
of pure and functionalized carboranethiols calculated using the same molecular 
orientations on Au(111) with (3$\times$3) and (6$\times$3) cell structures. The 
$z$-axis corresponds to the [111] direction and the $x$-axis is oriented along 
[101] direction. Work function values ($\Phi$ in eV) calculated for CT/Au(111) 
systems at full monolayer coverage. Charge transfers values ($\Delta Q$ in $e$) 
from the gold surface to individual molecules are given for their 
corresponding chemisorption geometries.}
\label{table4}\vskip1mm
\begin{tabular}{c|c|c|c|c|c} \hline
\hspace{3mm}\mbox{}Molecule\hspace{3mm}\mbox{} & 
\hspace{4mm}\mbox{}$\mu_{x}$\hspace{5mm}\mbox{} & 
\hspace{4mm}\mbox{}$\mu_{y}$\hspace{5mm}\mbox{} & 
\hspace{4mm}\mbox{}$\mu_{z}$\hspace{5mm}\mbox{} & 
\hspace{3mm}\mbox{}$\Phi$\hspace{3mm}\mbox{} & 
\hspace{3mm}\mbox{}$\Delta Q$\hspace{3mm}\mbox{} \\ \hline
\multicolumn{6}{c}{CT/Au(111)-(3$\times$3)} \\ \hline
M1 &-0.332 &-2.125 &~0.526 & 5.50 & 0.175\\
M9 &-0.641 &-1.333 &~4.270 & 4.96 &  0.096\\ \hline
M1-COOH &~1.945 &-0.949 &~0.204 & 5.57 & 0.160\\			
M9-COOH &~0.889 &-0.417 &~3.542 & 5.10 & 0.103\\ \hline
M1-NH$_2$ &~2.822 &-0.023 &-0.256  & 5.66 & 0.146\\             
M9-NH$_2$ &~2.551 &~1.424 &~3.220  & 5.17 & 0.083\\ \hline
\multicolumn{6}{c}{CT/Au(111)-(6$\times$3)} \\ \hline
M1 & \begin{tabular}{c} ~2.086 \\ -4.231 \end{tabular} & 
\begin{tabular}{c} ~0.357 \\ ~0.448 \end{tabular} &
\begin{tabular}{c} ~0.610\\ ~0.627 \end{tabular}  & 
5.52 & 
\begin{tabular}{c} 0.148 \\ 0.169 \end{tabular}\\[4mm]
M9 & \begin{tabular}{c} ~0.010 \\ -0.014 \end{tabular} & 
\begin{tabular}{c} -0.615 \\ ~3.188 \end{tabular} &
\begin{tabular}{c} ~4.421 \\ ~3.257 \end{tabular} & 
4.99 & 
\begin{tabular}{c} 0.109 \\ 0.097 \end{tabular}\\ \hline
M1-COOH  & \begin{tabular}{c} ~2.281 \\ -2.229 \end{tabular} &
\begin{tabular}{c} -1.220 \\ ~1.235 \end{tabular} & 
\begin{tabular}{c} -0.424 \\ ~0.248 \end{tabular} & 
5.58 & 
\begin{tabular}{c} 0.111 \\ 0.149 \end{tabular}\\[4mm]	
M9-COOH & \begin{tabular}{c} ~0.487 \\ ~0.089 \end{tabular} & 
\begin{tabular}{c} -1.440 \\ -1.440 \end{tabular} &
\begin{tabular}{c} ~3.358 \\ ~3.912 \end{tabular} & 
5.13 & 
\begin{tabular}{c} 0.052 \\ 0.113 \end{tabular}\\ \hline
M1-NH$_2$ & \begin{tabular}{c} ~2.448 \\ -2.617 \end{tabular} & 
\begin{tabular}{c} -1.468 \\ ~1.049 \end{tabular} &
\begin{tabular}{c} ~0.638 \\ -0.329 \end{tabular} & 
5.67 & 
\begin{tabular}{c} 0.150 \\ 0.165 \end{tabular}\\[4mm]            
M9-NH$_2$ & \begin{tabular}{c} ~2.582 \\ ~0.533 \end{tabular} & 
\begin{tabular}{c} -0.596 \\ -1.388 \end{tabular} &
\begin{tabular}{c} ~2.939 \\~4.507 \end{tabular} & 
5.15 & 
\begin{tabular}{c} 0.116 \\ 0.121 \end{tabular}\\ \hline
\end{tabular}
\end{table}

Instead of functional groups facing the same direction in the monolayer,
an alternating arrangement is possible where every two neighboring molecules 
face each other allowing the functional groups to form dimers. This type of 
intermolecular interaction corresponds to the (6$\times$3) unit cell with 
respect to the underlying gold substrate, as shown in Fig.~\ref{fig6}. The 
nearest-neighbor distance becomes 9.21 {\AA} between the molecules whose 
functional groups show dimer formation with increased steric demands while it 
is 7.95 {\AA} between molecules with non-dimering functional groups along the 
other lateral direction. Therefore, resulting molecular ordering breaks the 
uniformity of the adlayer morphology.   

On the (3$\times$3) structure, the shortest distance between the carboxyl 
proton to the boron vertex on the adjacent molecule in the lateral direction is 
2.59 {\AA} and 2.75 {\AA} for M1 and M9, respectively. Hence, COOH..BH hydrogen 
bonding is more favorable and has a larger effect on the adsorption 
characteristics. In the case of the amine group, these distances are 
3.49 {\AA} and 3.35 {\AA}, for M1 and M9 isomers, respectively.  On the 
(6$\times$3) structure, the protons mutually interact with the adjacent oxygens 
of the dimering carboxylic acids. The O-H distances are 1.56 {\AA} for 
M1-COOH and 1.55 {\AA} for M9-COOH. The separations between NH$_2$ dimers are 
3.06 {\AA} for M1-NH$_2$ and 3.42 {\AA} for M9-NH$_2$. Therefore, 
intermolecular interactions between the dimering molecules are more 
pronounced in the case of COOH-functionalized (6$\times$3) structures.   
In fact, COOH-dimer conformation gives the strongest binding among 
the other probable adsorption structures as seen in Table.~\ref{table3}.

Pure and functionalized carboranethiols chemisorp at the 3-fold hollow 
site both on (3$\times$3) and (6$\times$3) unit cells in agreement with 
experiments.\cite{Hohman2009,Goronzy}. Isolated molecules, on the other hand, 
tend to be adsorbed closer to the bridge site. As a common feature for 
all coverage models, the S atom lifts one of the 3-fold coordinated gold 
atoms from surface plane leading to a significantly shorter S-Au bond relative 
to the other two. This causes a local distortion around the chemisorption site 
which is particularly more pronounced in the densely packed arrangements 
relative to that of the isolated molecules. The S-Au and S-Cage bonds reflect 
typical distances similar to the isolated cases as seen in Table~\ref{table3}. 
As a result, the thiol plays an important role as an anchor between the gold 
surface and the carborane monolayer. Furthermore, neither attachment of the 
functional groups nor increased molecular density on the gold surface appear to 
noticeably influence the skeletal stability of the molecules.  

The tilting angles are expected to be reduced due to both increased molecular 
density in the monolayers and steric preferences of functionalized CTs. Indeed, 
a decrease of the tilting angles is seen for molecules leaning toward 
their nearest-neigbors forming the (3$\times$3) structure. Particularly, the 
comparison between the (5$\times$5) and (3$\times$3) cases shows that the most 
significant reduction in the tilting angles happens for the pure and 
functionalized M9 isomers due mainly to stronger intermolecular forces. 
The interactions between the M1 isomers are weaker, because they have 
considerably smaller dipole moments with orientations almost parallel to the 
surface as presented in Table~\ref{table4}. Contrary to (3$\times$3) cases, 
the molecules in dimer formation on (6$\times$3) cell get even more tilted 
relative to isolated molecules on the (5$\times$5) structure. In the case of 
pure isomers, molecules are not tilted toward each other. They become 
oppositely tilted. On the other hand, dimering carboxyl and amine 
groups cause the CTs to lean toward each other due to hydrogen bonding 
between them. The (3$\times$3) and (6$\times$3) conformations allow
different hydrogen bonding possibilities between the functionalized CTs. The 
dimer conformation in which more positively charged part of a functional group 
facing more negatively charged part of the adjacent group at a distance 
of 1.54 {\AA} builds an attractive force between the nearest-neighbor 
molecules, allowing them to lean more towards each other on the surface. 

As going from isolated to full monolayer coverages, calculated 
dissociative chemisorption energies show stronger binding between the CT 
molecules and the gold surface due to intermolecular interactions in 
monolayers. Furthermore, the functional groups have significant influence on 
the binding energies. In particular, carboxyl acid group appears to be 
more effective to get more stable monolayers relative to the amine group. 
This is particularly noticeable in the COOH attached carborane coating on the 
(6$\times$3) cell of Au(111). The chemisorption energies of M1 and M9 isomers 
with the same functional group become almost indistinguishable, which 
allows a possibility of mixed monolayer realization. 

\begin{figure*}[htb]
\includegraphics[width=16.3cm]{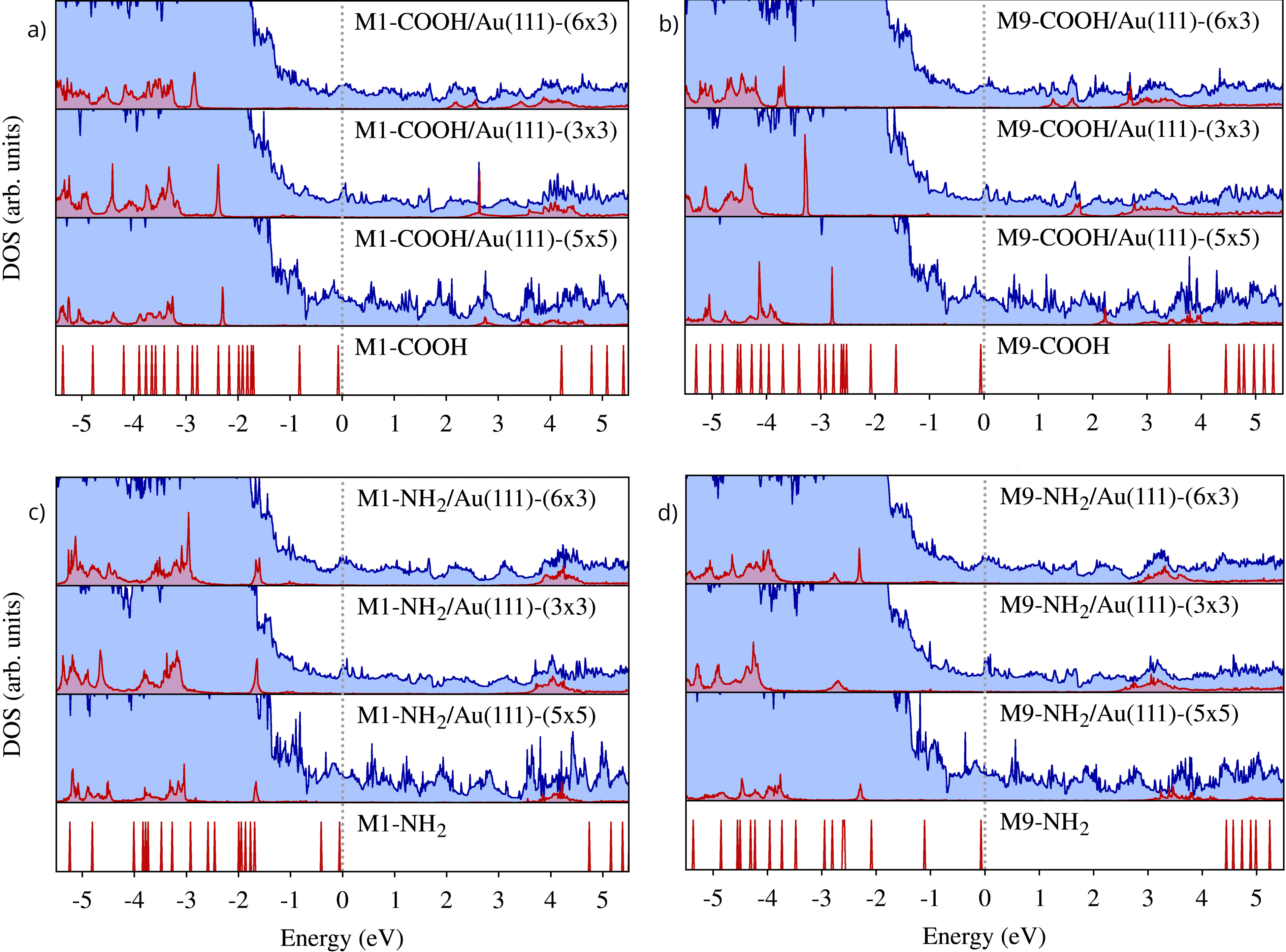}
\caption{Partial densities of states for a) M1COOH, b) M9COOH, c) M1NH$_2$, 
and d) M9NH$_2$ on Au(111) with (5$\times$5), (3$\times$3), and 
(6$\times$3) structures calculated using the SCAN+rVV10 functional. The bottom 
panels show the energy levels of the corresponding isomers in the gas phase. 
Vertical dashed lines indicate the Fermi energy for each system.}
\label{fig7}
\end{figure*}

The charge transfer from the gold surface to CTs is mainly characterized by 
positional isomerization which modifies molecular dipole moments. The amount 
of charge transfer for M9 derivatives is slightly smaller than that of M1. 
In addition, as the molecular density increases from isolated to full monolayer 
coverage, a noticeable increase is seen in the amount of charge transfer per 
molecule. Such an increasing charge transfer corresponds to a stronger 
binding in favor of dense packing. Moreover, the functional groups appear to 
have a secondary effect on the charge transfer values. Molecular dipoles are 
useful in modifying the work function of gold. Kim~\textit{et al.} reported 
that the work function of gold decreases with M9 coverage and increases with M1 
adsorption. A similar trend in the calculated work function values is seen in 
Table~\ref{table2} and in Table~\ref{table4}. In consistency with experimental 
measurements, a monolayer of M9 decreases the work function by 0.45 eV 
relative to that of the bare Au(111) while a monolayer of M1 increases it by 
0.09 eV in the case of (3$\times$3) cell. Similar results were obtained 
for carboranethiol SAMs with the (6$\times$3) cell. Experiments demonstrated 
tunability of the work function by depositing mixed SAMs with varying 
concentrations of M1 and M9 isomers.\cite{Kim2014} In addition, a comparison 
between the isolated and monolayer cases indicate that increasing molecular 
density reduces the the work function of gold. The reduction is $\sim$0.2 eV and 
$\sim$0.4 eV for M1 and M9 isomers. Both carboxylic acid and amine functional 
groups are effective in modifying the work function relative to pure M1 and M9 
monolayers on gold. These groups not only offer a modifiability of the work 
function over a wider range but also play an important role in the 
formation of well-ordered and functional SAMs for various applications.

\subsection{Electronic Structures of Carboranethiols on Au(111)}

The electronic densities of states (DOS) with partial projections were 
calculated for isolated and monolayer CTs on unreconstructed Au(111). 
The M1 and M9 isomers with amine and carboxyl functional groups considered as 
presented in Fig.~\ref{fig7}. The DOS structures for different unit cells were 
aligned with respect to deep core states. In the gas phase, M1-NH$_2$ and 
M9-NH$_2$ have HOMO-LUMO separations of 4.7 eV and 4.4 eV, respectively.While 
the separation between the frontier orbitals is 4.2 eV for M1-COOH, it is 
significantly smaller for M9-COOH as being 3.3 eV.  

Chemisorption of amine and carboxyl functionalized CTs on Au(111) has a drastic 
influence on the frontier molecular orbitals. The partial DOS (PDOS) 
structures show that the HOMO levels strongly resonate over a wide range of the 
occupied gold states as a result of the bond formation between the thiol 
terminal and the surface gold atoms. Upon adsorption, the HOMO-1 and LUMO 
states broaden and shift down more than 1 eV to lower energies. The shift 
increases going from the isolated to the monolayer phases, especially, in favor 
of the dimer conformation on the (6$\times$3) cell in consistency with the 
calculated chemisorption energies. Moreover, this effect for both occupied and 
unoccupied DOS features is particularly more pronounced in the carboxylated CT 
cases. 

The main PDOS features of amine functionalized M1 and M9 molecules are 
essentially kept as going from isolated (5$\times$5) to monolayer (3$\times$3) 
and (6$\times$3) coverages. In particular, for M1-NH$_2$, the HOMO-1 at around 
-2 eV, a satellite of occupied molecular states in the interval between -3 eV 
and -5.5 eV, and a group of antibonding molecular states centered around 4 eV, 
basically keep their positions and forms throughout different coverages and 
conformations. Slight changes between the PDOS of NH$_2$ functionalized isomers 
on gold can be attributed to relatively weaker intermolecular interactions. In 
the case of carboxyl functionalized CTs on the gold surface, the shifting and 
broadening of both occupied and unoccupied energy states associated with the 
CTs increase depending on the molecular density. Dense packing of COOH 
functionalized CTs gives rise to hydrogen bonding and favorable dipole-dipole 
interactions between neighboring molecules within the monolayers.

\section{Conclusions}

The functional groups bring anisotropy to CTs and increase their steric 
requirements depending on positional isomerization. The DFT calculations 
including vdW effects reveal that thiol heads form an inequivalent 3-fold 
coordination with the gold atoms. Isolated molecules relax closer to bridge 
site while densely packed molecules prefer the hollow site. One of the surface 
gold atoms raised up from the surface plane by the thiol terminal causing a 
tilted adsorption. The tilt relative to the surface normal and the absolute 
value of chemisorption energies increase with the functional groups.    
The change in the chemisorption energies favors high molecular density. 
Therefore, the DFT calculations indicate that functionalization of CT isomers 
lead to more stable monolayers on the gold surface. Carboxylic acid group is 
slightly more effective than amine group, in increasing the absolute value of 
dissociative chemisorption energies. 

In addition to the cage geometry of carboranethiols, adsorption geometries 
and energies are mainly characterized by the bonding between the thiol and 
surface gold atoms, the molecular dipole moments and the functional groups. 
The results clearly show the effect of intermolecular interactions which are 
effective in the densely packed arrangements. Presence of functional groups 
make two different conformations probable for monolayer structures. The 
functional groups align in the same lateral direction or form dimers facing 
each other, which are commensurate with the (3$\times$3) and (6$\times$3) cells 
with respect to underlying Au(111). These dense packing conformations give rise 
to hydrogen bonding and favorable dipole-dipole interactions in monolayers.

The functional groups in conjunction with positional isomerization influence 
the molecular dipole moments. A dipole along surface normal decreases 
the work function while a dipole parallel to increases it. The functional 
groups, which are exposed to the environment from the SAM surfaces, have 
different electrochemical responses allowing various designs for applications. 
Therefore, the carboxyl and amine groups are useful in functionalization of SAM 
structures and still offer tunability of the work function with desirable 
properties.

Although amine and carboxyl groups have different electronic charge states one 
being an electron donor and the other being an electron acceptor, they both
make CT molecules to bind to the gold surface stronger. More studies are needed 
to explore the effect of other possible functional groups on the characteristics 
of SAMs.

\section*{Acknowledgements}
\noindent This study was supported by Tubitak, The Scientific and Technological 
Research Council of Turkey (Grant No: 116F174) and Bal{\i}kesir University 
(Project No: BAP 2018/039). The calculations that conducted in this article have 
been performed on the High Performance and Grid Computing Center (TRUBA parallel 
computer center).

\bibliography{Refs} 

\end{document}